\documentclass[fleqn,10pt]{wlscirep}
\usepackage[utf8]{inputenc}
\usepackage[T1]{fontenc}
\usepackage{bm}
\usepackage{graphicx}

\title{Revealing a full quantum ladder by nonlinear spectroscopy}

\author[1,*]{Darius Abramavicius}
\affil[1]{Institute of Chemical Physics, Physics Faculty, Vilnius University,
Sauletekio 9-III, LT-10222 Vilnius Lithuania}

\affil[*]{darius.abramavicius@ff.vu.lt}

\begin{abstract}
Coherent two-dimensional spectroscopy in IR or visible region is very
effective for studying correlations, energy relaxation/transfer pathways
in complex multi-chromophore or multi-mode systems. However it is
usually restricted up to two-quanta excitations and their properties.
In this paper an arbitrary level of excitation is suggested as the
utility to scan nonlinear potential surfaces of quantum systems up
to a desired excitation degree. This can be achieved by a simple three-pulse
laser spectroscopy approach. Accurate evaluation of high-level anharmonicities
as well as transition amplitudes can be directly obtained. Additionally,
questions regarding the quantum nature of the probed system can be
addressed by studying absolute peak positions.
\end{abstract}
\begin{document}

\flushbottom
\maketitle
%
%
\thispagestyle{empty}


\section*{Introduction}

Impulsive laser spectroscopy is one of the main material science utilities
for revealing excited states dynamics. The forefront in the nonlinear
domain, the two-dimensional coherent spectroscopy (2DIR in IR region,
2DES in visible region) \cite{chemrev09,Zhuang2009}, which has been
applied in solid \cite{Cho2009,Wen2013}, protein \cite{Schlau-Cohen2011,Reppert2016},
liquid \cite{Hume2019,Thamer2015} as well as gaseous \cite{Bruder2018}
materials. It has been also utilized in microscope configuration \cite{Tiwari2018}.
2D spectroscopy evolved from of the transient absorption spectroscopy.
The two-pulse pump probe or transient absorption approach provides
the time resolution by measuring the transient evolution of difference
absorption induced by the excitation pump pulse and the detection
probe pulse. Since the two pulses provide the time resolution as the
delay between pump and probe pulses, the measurement essentially reflects
the change of induced response of the sample in time. By adding a
third laser pulse in e. g. in a three pulse photon echo peak shift
(3PEPS) approach, the additional time delay of intraband coherence
allows to additionally follow the environment-induced correlation
functions of the energy gap fluctuations \cite{WimP.deBoeij1996,Everitt2001,Gibson2009}.
The pump probe and 3PEPS rely on the four wave mixing (4WM) process
in the nonlinear system, however, from a theory point of view, the
specific convolutions of the third order molecular response function
are performed and detected in these approaches \cite{WimP.deBoeij1996,mukamelbook}.
The convolutions are lifted by ultrashort excitation pulses imitating
elementary perturbative excitations. A natural upgrade to the 4WM
techniques then results in four pulse setup when the full deconvoluted
information contained in the third order response function is retrieved
by the two-dimensional heterodyne detected coherent spectroscopy.
Denoting $i-$th pulse wavevector by $k_{i}$, in this case, the specific
component of the four wave mixing process can be associated with the
system-field interaction configuration, and by collecting signals
$-k_{1}+k_{2}+k_{3}$, $k_{1}-k_{2}+k_{3}$, $k_{1}+k_{2}-k_{3}$,
all available information contained in the third order response function
with largest possible time and frequency resolution can be gathered
\cite{chemrev09,Zhuang2009}. Ideally since each excitation pulse
provides a single (one quantum) excitation with a dedicated wave vector,
the specific peaks in spectra can be associated with a minimal set
of interaction pathways. The obtained two-dimensional (2D) spectrum
sometimes is denoted as representing a picture of \textquotedbl quantum
tomography\textquotedbl{} \cite{Yuen-Zhou2011,Abramavicius2010}.

In the visible optical regime the 4WM measurements mainly demonstrate
the properties of the single exciton manifold in $-k_{1}+k_{2}+k_{3}$
and $k_{1}-k_{2}+k_{3}$ configurations. The single excitation manifold
is very important in physical chemistry as the single excitations
are main players in excitation transport, charge separation, charge
transport processes in molecular and biological functional systems
\cite{Collini2010,Meneghin2018,Palecek2017,Scholes2017,C7CP01673E,Fuller2014,Lim2015,Thyrhaug2018}.

The double excitation manifold, i. e. the set of states, which are
achieved by sequential two photon absorption participate in excited
state absorption in transient absorption, 3PEPS and 2DES. The double-exciton
level structure or exciton-exciton correlation is directly achieved
in $k_{1}+k_{2}-k_{3}$ (double quantum coherence) detection \cite{Abramavicius2012,Kim2009}
or in fifth order 2D spectra where exciton-exciton annihilation becomes
considerable \cite{Dostal2018,Bruggemann2011,Bubilaitis2019,Suz2019}.

While the dynamical aspects of energy migration is the main focus
in electronic spectroscopy, the 2D spectroscopy approach in IR regime
\cite{Fayer2013} has been utilized for molecular structure determination
\cite{Zhuang2009,Remorino2012,Smith2010} and for tracking thermal
fluctuations and chemical reaction dynamics \cite{Baiz2010,Fayer2009,Ji2010,NgPack2019}
via vibrational resonances. Vibrational nonlinear spectroscopy allows
direct measurement of diagonal and off diagonal vibrational anharmonicities
\cite{Cyran2015,Guo2015}. These parameters are related to single
and double quanta properties, hence, denote the lowest order nonlinear
properties of the complex vibrational potential surfaces \cite{Laane2009,Laane2000,Chen2015}.
Several spectroscopic methods, including infrared and ultraviolet
absorption, Raman, jet-cooled laser-induced fluorescence, have been
utilized to map out the whole series of vibrational quantum states
of molecules in their ground and excited electronic states \cite{Knochenmuss2018,Zuniga2012,Dong2010}.
These multiple quanta properties are very important for numerous molecular
processes such as chemical reactions \cite{Yan2014,Skulason2010},
isomerization or inversion \cite{Pechukas1981,Satzger2004,Johnson2015,Balevicius2019,Hauer2013},
as well as energy transfer during inelastic collisions proceed along
vibrational pathways that are governed by vibrational potential energy
surfaces \cite{Jambrina2019,Saez-Rabanos2019}. Development of spectroscopic
approaches to reveal the whole vibrational (or electronic) potential
surface is thus highly demanded and simple nonlinear spectroscopy
probes can aid in this direction.

In this letter a set of nonlinear optical measurements is suggested
to directly measure multiple-quanta up to a desired degree excitation
properties. The obtained two-dimensional spectra then can be employed
to reveal the energy level structure of a potential surface what is
necessary for reconstruction of molecular potential surfaces. A side
question regarding the quantum nature of vibrations can be addressed
as well.

\section*{Theory and simulations}

Optical excitations of a quantum system can be easily described by
the response function theory at various orders with respect to the
system-field interactions. Using a response function formalism the
induced polarization's $j$-th order to the field component is given
by
\begin{eqnarray}
P^{(j)}(t) & = & \int_{0}^{\infty}dt_{j}\dots\int_{0}^{\infty}dt_{1}\ S^{(j)}(t_{j}\dots t_{1})\times\nonumber \\
 &  & \ \ E(t-t_{j})E(t-t_{j}-t_{j-1})...E(t-t_{j}-...t_{1})\label{eq:polarization-general}
\end{eqnarray}
 In an electric dipole approximation all even orders lead to vanishing
response of bulk samples. We next consider a single color source excitation
field with frequency resonant to one particular anharmonic molecular
mode. Before hitting the sample, the field is split into a set of
excitation fields
\begin{eqnarray*}
E(t) & = & \sum_{i}^{N}E_{i}(t)=\sum_{i}^{N}A_{i}(t)\exp(i\bm{k}_{i}\bm{r}-i\omega t)+c.c.
\end{eqnarray*}
 Here $A_{i}(t)$ is a slowly varying amplitude of the $i$-th component
whose wavevector is $\bm{k}_{i}$. In nonlinear regime the phase matching
phenomenon is an additional necessary companion in the optical spectroscopy.
Indeed, multiplication of such fields in Eq. \ref{eq:polarization-general}
results in superpositions of wavevectors, and thus the signals are
being generated with the specific new signal wave vectors given by
$\bm{k}_{s}=n_{1}\bm{k}_{1}+n_{2}\bm{k}_{2}+...n_{N}\bm{k}_{N}.$
Here $N$ is the number of incoming fields, $n_{1},...,n_{N}$ are
integers each taking values in the interval $-j,-j+1...j-1,j$, where
$j$ is the order of nonlinearity and we have additionally $|n_{1}|+|n_{2}|+...+|n_{N}|=j$.

\begin{figure}
\begin{centering}
\includegraphics[width=0.5\textwidth]{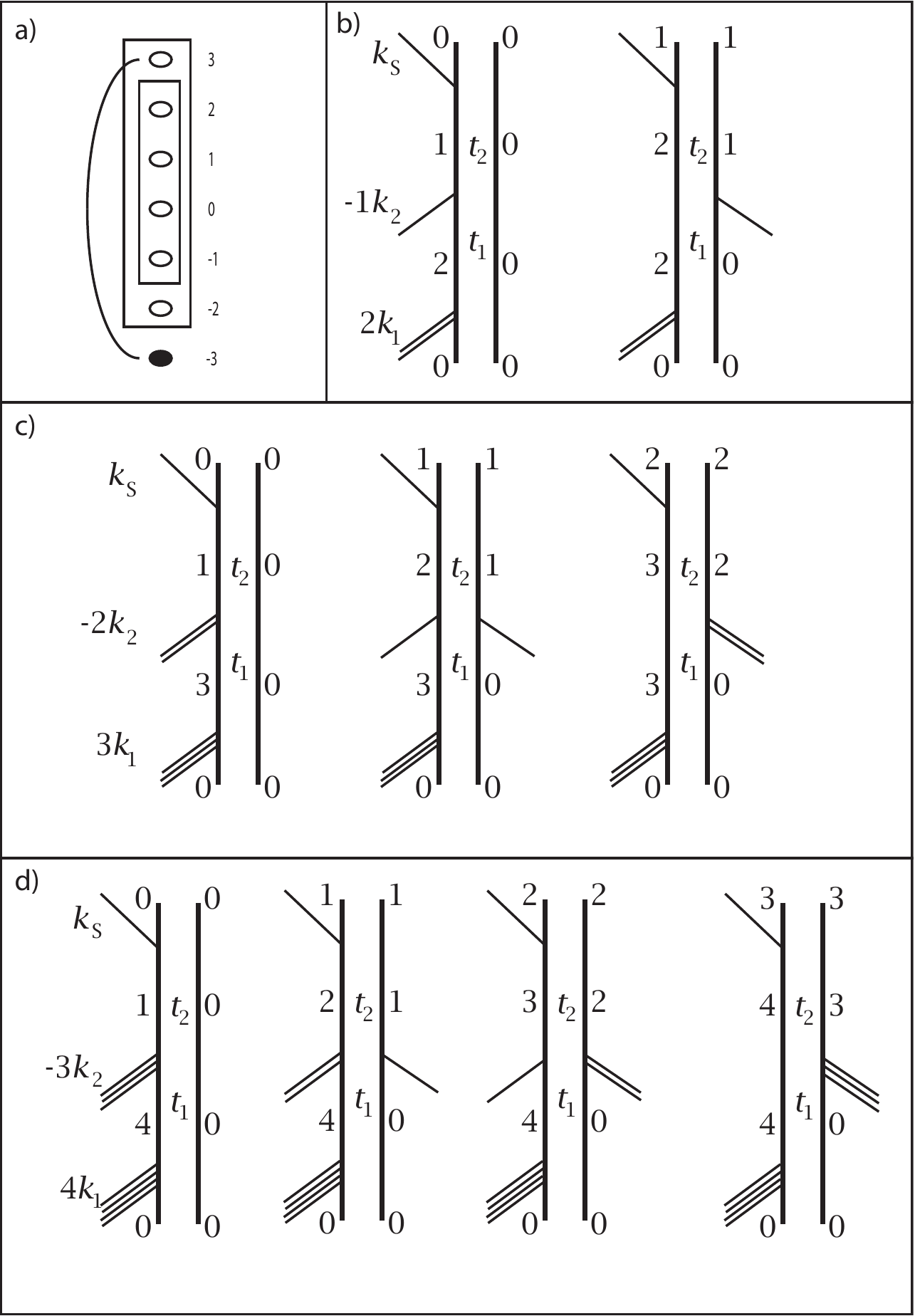}
\par\end{centering}
\caption{\label{fig:Diagrams}a) Resolution of wave vector components in phase
matching process as viewed from spatial Fourier transform. Feynman
diagrams for the $k_{S}=nk_{1}-(n-1)k_{2}$ phase matching configuration
with two excitation pulses: b) at 3-rd ($n=2$), c), at 5-th ($n=3$),
d) and at 7-th ($n=4$) order to the field.}
\end{figure}

Spatial detuning of laser rays can be used to reveal specific output
wavevectors. The collinear geometry can be also used to increase the
field overlap region and specific phase matched configuration can
be obtained by phase cycling. Phase cycling for $N$ incoming pulses
is formally equivalent to the spatial Fourier transformation in $N$
dimensions. To isolate an $m$-th order component ($m+1$ wave mixing)
on a single dimension one needs to use at least $m+1$ equally distributed
phase values as dictated by periodicity of Fourier transform and Nyquist
theorem (see Fig. \ref{fig:Diagrams} a). Indeed the discrete transform
of equally spaced $M$ samples in $x$ dimension of a function $f(x_{i})=f_{i}$
yields $M$ samples in conjugate $k_{x}$ dimension. So for distinguishing
$\pm1$ output values one needs to provide $M=4$ input distinct values
corresponding to $-1\Delta_{x},0\Delta_{x},+1\Delta_{x},+2\Delta_{x}$
or $-1k_{x},0k_{x},+1k_{x},+2k_{x}$ samples. Notice that in $M=4$
case due to periodicity of Fourier transform $-2k_{x}\equiv+2k_{x}$
cannot be distinguished.

At first, we may consider designing a high order multidimensional
spectroscopy to probe highly lying excited states and reveal complete
nonlinear response function in the spirit of 2DES. Notice that at
3-rd order to the field by adding the third harmonic generation configuration,
$k_{s}=k_{1}+k_{2}+k_{3}$, the number of independent phase matching
components is $2^{3}/2=4$ (half of $2^{3}$ components are conjugate
to each other), at $j$-th order we thus have $2^{j}/2$. Hence, in
order to gather the complete information available at $j$-th order,
the number of phase matched configurations which have to be inspected
scales exponentially with the order in the field.

Such ``complete-information'' way of thinking calls for an experiment
where for $j-$th order we should consider $j$ laser pulses. However,
it becomes impossible to continue in this spirit as the dimensionality
of the phase matching configurations and dimensionality of signal
as a function of time delays, grows up and the amount of gathered
data becomes prohibitively large.

Hence, as higher-energy and quanta states become populated in higher-order
spectroscopies the focus of the problems must shift towards the multiple-excitation
characteristics. The most important information then for example is
the energy level configuration or in other words the shape of potential
surface \cite{LeRoy2006}. It should be noted that this used to be
the main focus of the 2DIR where vibrational anharmonicities have
been resolved\cite{Fayer2013}.

Notice that high order nonlinear interaction process is not necessarily
related to a large number of laser pulses. Indeed a single field induces
response at an arbitrary order to the field and the phase matching
unravels specific interaction pathways. The ideal approach is to use
a smallest number of excitation pulses to reveal the desired information
on $n$-quanta energy levels. A sequential process may be designed
by two pulses to drive the excitation up and down the ladder to the
quantum states and then detect the optical coherence field.

Consider for example a response induced by two incoming laser pulses.
The first pulse corresponds to wavevector $k_{1}$ while the second
to $k_{2}$. Let's inspect the induced polarization detected at $k_{S}=nk_{1}-(n-1)k_{2}$,
where $n$ is an arbitrary integer. The Feynman diagrams corresponding
to such process are given in Figure \ref{fig:Diagrams}b-d. In this
setup, during the first delay interval $t_{1}$ the state of the system
is given by coherent superposition of the ground state and the $n$-th
excited state that is reached by absorption of $n$ photons. In the
case of small anharmonicity, the corresponding density matrix element
oscillates with the frequency approximately equal to $n\omega_{0}$,
where $\omega_{0}$ is the frequency of the photon, provided anharmonicity
is small. The following $n-1$ interactions (either on the left or
on the right of the diagram) with the second pulse after delay $t_{1}$
generate interband coherences that range from $|1\rangle\langle0|$
to $|n\rangle\langle n-1|$. The diagram thus projects information
about $1,2,...,n$ quanta excited states onto a two-dimensional function
of delays $t_{1}$ and $t_{2}$. The latter can be Fourier transformed
into a two-dimensional spectrum $S(\omega_{1},\omega_{2})$.

Notice that considering $n$ quanta excitations the strongest transitions
are between $n\to n\pm1$ levels; these transitions originate from
a single light quantum absorption/emission. Hence, in $k_{S}=nk_{1}-(n-1)k_{2}$
configuration all involved transitions are of single quantum absorption/emission,
so the signal intensity is never scaled by a small anharmonic correction
of the transition dipole.

As an example the multiple quanta spectra of a Morse oscillator is
presented in the following. The Morse oscillator is a unique anharmonic
quantum system whose energy levels and transition amplitudes are available
analytically \cite{Lima2005}. The potential surface of the oscillator
can be given by
\begin{equation}
V(x)=D(1-\exp(-\alpha x))^{2}
\end{equation}
 with $D$ being the classical ionization energy, $\alpha$ is the
curvature parameter of the potential valley. For the case when mass
of the oscillator is 1/2, the oscillator fundamental frequency is
equal to $\omega_{\alpha}=2\alpha\sqrt{D}$. By taking for convenience
$\alpha=1$ we find that $D$ is the only anharmonicity parameter.
Denoting $D=(\mathcal{N}+1/2)^{2}$, the integer part of $\mathcal{N}+1/2$
(which we denote by $N=\mathrm{Int}\left[\mathcal{N}+1/2\right]$)
equals to the number of discrete quantum levels. Essentially the system
is more anharmonic as the number of discrete energy levels gets smaller
and the oscillator turns into harmonic system with $D\to\infty$.
Energy spectrum of the discrete states of the oscillator in this case
is
\begin{equation}
E_{n}=-(\mathcal{N}-n)^{2}
\end{equation}
 with $n=0$,1 ... $N-1$. In the following we assume that the dissociation
is not relevant and we do not consider the continuum part of the spectrum,
since its contribution to the spectrum is quite week (checked numerically,
not shown).

The next required quantity is the optical transition dipole moment.
In order to determine it, it is necessary to define the polarization
operator. The polarization operator for a harmonic oscillator is often
assumed to be proportional to the coordinate operator, $\hat{P}=\mu\hat{x}$.
We next assume that the optical field only involves transitions between
adjacent energy levels and other transitions are off resonant. We
thus consider only matrix elements $\langle n|\hat{P}|n+1\rangle$
\cite{Lima2005}:
\begin{equation}
\mu_{n}\equiv\mu\langle n|\hat{x}|n+1\rangle=\mu\frac{\sqrt{\left(\mathcal{N}-n\right)\left(\mathcal{N}-n-1\right)}}{\mathcal{N}-n-1/2}\left[\frac{\left(n+1\right)}{\left(2\mathcal{N}-n\right)}\right]^{\frac{1}{2}}
\end{equation}
which for large $\mathcal{N}\gg n$ reduces to the result of a harmonic
oscillator $\mu_{n+1}/\mu_{n}\approx\sqrt{\left(n+2\right)/\left(n+1\right)}$.

Consequently, we can write analytically the resulting expressions
for the 2D peak intensities. For example for $k_{s}=4k_{1}-3k_{2}$
we have the set of peaks whose amplitudes in the harmonic case are
approximately equal to $4!\mu_{0}$ since the system in each diagram
absorbs and emits up to 4 quanta. In general for the signal component
$k_{S}=nk_{1}-(n-1)k_{2}$ the peak intensities will be $n!\mu_{0}$.

All distinct diagrams have positions determined by resonant transition
energies. The fundamental energy gap
\begin{equation}
\omega_{0}=E_{1}-E_{0}=2\mathcal{N}-1,
\end{equation}
and correspondingly the sequence of energy gaps
\begin{equation}
E_{n+1}-E_{n}=\omega_{0}-2n
\end{equation}
 is linear with $n$. This sequence is directly observed in $k_{S}=nk_{1}-(n-1)k_{2}$
configuration.

\begin{figure}
\begin{centering}
\includegraphics[width=0.5\textwidth]{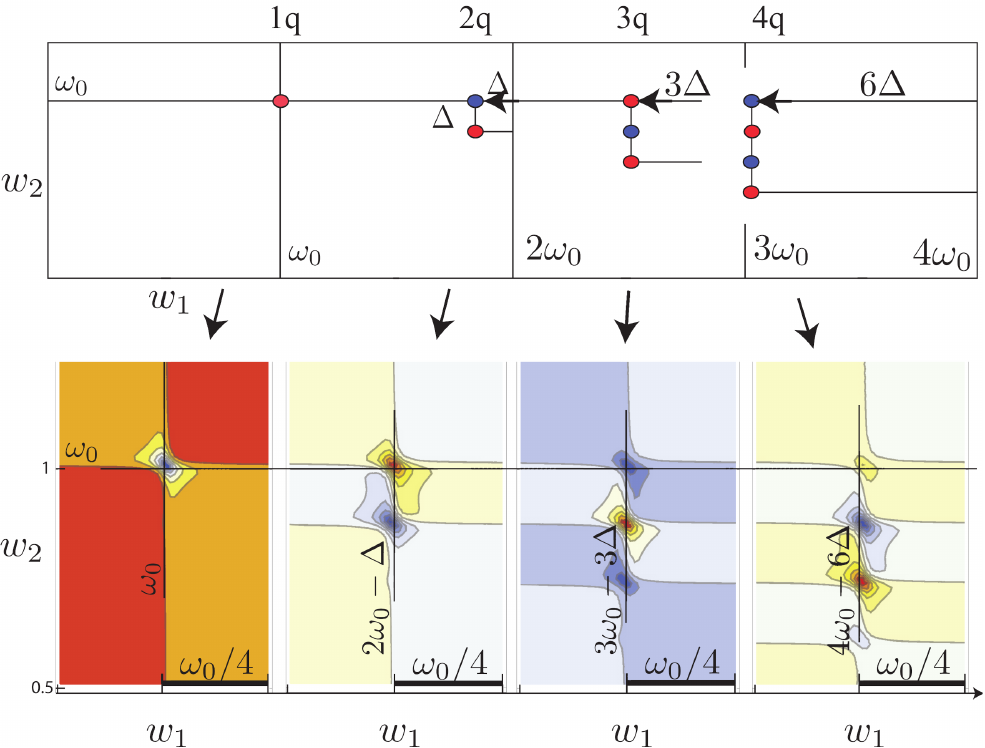}
\par\end{centering}
\caption{\label{fig:Diagrams-1}Top: scheme of peaks in the $k_{S}=nk_{1}-(n-1)k_{2}$
measurement of Morse oscillator corresponding to (from left to right)
$n=1$, $n=2$, $n=3$ and $n=4$. $\omega_{0}$ is the fundamental
frequency and $\Delta$ is the lowest anharmonicity. Bottom - the
spectrum of the Morse oscillator obtained from solving Liouville equation
and the signal (Eq. \ref{eq:response-quantum}) is reconstructed by
phase cycling. The damping is included as an additional factor to
the signal as $\exp(-\gamma(t_{1}+t_{2}))$ with $\gamma=0.02\omega_{0}$
so that all resonances are explicitly expressed.}
\end{figure}

By using the diagram approach we can easily show what peaks will be
visible in 2D spectrum for various wavevector configurations for Morse
oscillator. The oscillation frequency in the $t_{1}$ interval is
\begin{equation}
E_{n}-E_{0}=n\omega_{0}-\Delta_{n}
\end{equation}
with $\Delta_{n}=n(n-1)$, so the sets of peaks in $k_{S}=nk_{1}-(n-1)k_{2}$
measurement will appear at frequency $w_{1}=n\omega_{0}-\Delta_{n}$.
During $t_{2}$ the system will oscillate with frequencies corresponding
to energy gaps $E_{i+1}-E_{i}=\omega_{0}-\Delta_{2}i$ for $i=0,1...n$.
Such situation is demonstrated on the top of Figure \ref{fig:Diagrams-1}.

To demonstrate the above described approach in more realistic scenario,
we next present numerical simulations of the excitation and detection
process of the Morse oscillator with dephasing and lineshapes. We
set the oscillator parameter $\mathcal{N}=$8, so that the number
of bound states is $N=8$. The initial state of the system is equal
to the ground state $|0\rangle$. For the excitation process, which
we represent as action by nonlinear excitation operator, $X_{\phi}\bm{\rho}_{0}$,
we assume excitation by an optical field resonant to the fundamental
frequency and with the bandwidth covering all transitions $n\to n+1$.
In order to realize that the excitation/deexcitation is nonperturbative
(as in real experiment) we numerically integrate the interaction picture
Schr\"{o}dinger equation with the excitation field turned on. For Liouville
equation this results in iteration of equation
\begin{equation}
\dot{\bm{\rho}}=-\mathrm{i}\left(\bm{\mu}_{\phi}\bm{\rho}-\bm{\rho}\bm{\mu}_{\phi}\right),
\end{equation}
 here
\begin{equation}
\bm{\mu}_{\phi}=A\left(\begin{array}{ccccc}
0 & \mu_{1}e^{-i\phi} & 0 & ... & 0\\
\mu_{1}e^{i\phi} & 0 & \mu_{2}e^{-i\phi} & ... & 0\\
0 & \mu_{2}e^{i\phi} & 0 & ... & 0\\
... & ... & ... & ... & ...\\
0 & 0 & 0 & ... & 0
\end{array}\right)
\end{equation}
 is the transition dipole operator with only RWA terms including the
excitation intensity $A$ and phase $\phi$ (the phase is necessary
for phase cycling to recover correct phase matching contributions).
This equation is propagated for an infinitesimal time interval $\delta t\to0$
with the given intensity $A$, so that $A\delta t$ is not vanishing
and can be tuned to populate the chosen number of states.

So the protocol is as follows. We take the initial density matrix
\begin{equation}
\bm{\rho}_{0}=\left(\begin{array}{ccc}
1 & 0 & ...\\
0 & 0 & ...\\
... & ... & ...
\end{array}\right),
\end{equation}
perform excitation $X_{\phi_{1}}\bm{\rho}_{0}$ with phase $\phi_{1}$,
then we perform the density matrix propagation according to Liouville
equation with the field switched off to obtain $\mathcal{G}(t_{1})X_{\phi_{1}}\bm{\rho}_{0}$.
The second excitation follows with the phase $\phi_{2}$: $X_{\phi_{2}}\mathcal{G}(t_{1})X_{\phi_{1}}\bm{\rho}_{0}$,
propagation with the field off for time delay $t_{2}$: $\mathcal{G}(t_{2})X_{\phi_{2}}\mathcal{G}(t_{1})X_{\phi_{1}}\bm{\rho}_{0}$,
and finally we perform the final detection to obtain the signal
\begin{equation}
S(t_{1},t_{2})=\mathrm{Tr}\{\bm{\mu}_{L}\cdot\mathcal{G}(t_{2})X_{\phi_{2}}\mathcal{G}(t_{1})X_{\phi_{1}}\bm{\rho}_{0}\}.\label{eq:response-quantum}
\end{equation}

We emphasize the difference between this expression and standard response
function theory, where each excitation event is a linear operation.
In our approach we assume that each operation $X_{\phi_{1}}\bm{\rho}$
is nonlinear and thus generates the whole spectrum of harmonics.

On the bottom of Fig. \ref{fig:Diagrams-1} we present a series of
2D spectra after Fourier transform $S(t_{1},t_{2})\to S(\omega_{1},\omega_{2})$
for the signal $k_{S}=nk_{1}-(n-1)k_{2}$ with $n=1,2,3,4$ that correspond
to the nonlinear signals at $1,3,5,7$ order. The line broadening
was kept small.

\begin{figure}
\begin{centering}
\includegraphics[width=0.5\textwidth]{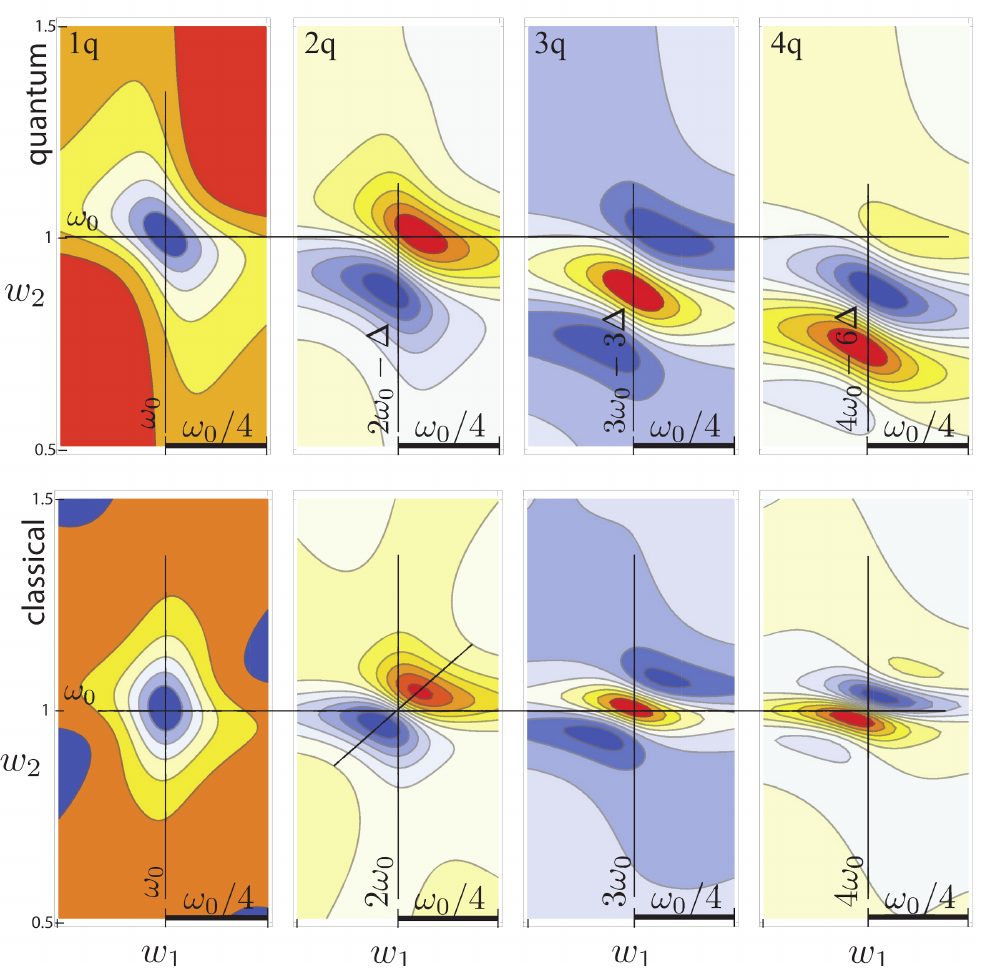}
\par\end{centering}
\caption{\label{fig:quantum-broad-classical}Numerical simulations of $k_{S}=nk_{1}-(n-1)k_{2}$
signals of Morse oscillator corresponding to (from left to right)
$n=1$, $n=2$, $n=3$ and $n=4$ cases for the quantum (top) and
classical (bottom) oscillators. $\omega_{0}$ is the fundamental frequency
and $\Delta$ is the lowest anharmonicity. Line broadening parameter
is set to $\gamma=0.1\omega_{0}$.}
\end{figure}

The common critique to these types of simulations is that the line
broadening has to be maintained small in order to observe the peak
pattern and to determine the anharmonicity. However, in this type
of signal that is not an issue because we observe not a single pair
of peaks, but the whole train of peaks. The same simulations of 2D
spectra with large line broadening ($\gamma=0.1\omega_{0}$) is shown
in Figure \ref{fig:quantum-broad-classical}. On the top row we can
observe that overlapping peaks become squeezed, but the number of
peaks and peak splitting can be uniquely determined.

--------

Additional question on the quantum nature of oscillations can be addressed
in this case. We additionally show the corresponding classical simulation
results, i e. for the same type of Morse potential but obtained from
purely classical equations of motion. We find that the peak pattern
is quite similar, however its overall absolute position becomes shifted
2 the series of peaks becomes positioned symmetrically with respect
to theoretical central point equal to $(w_{2},w_{1})=(\omega_{0},n\omega_{0})$,
while the series is shifted in the case of quantum oscillator.

\section*{Discussion and conclusions}

So where is the next level of impulsive spectroscopy? In this paper
the direction towards higher order nonlinear interactions is considered.
The fifth order coherent spectroscopy has been considered in IR and
visible and the complex information on the correlations in quantum
systems has been represented in 3DIR \cite{Borek2014} or 3DES \cite{Fidler2010,Zhang2015,Li2013} or 3D fluorescence \cite{Mueller2019}.
By spreading frequencies into a
third spectral dimension high
frequency vibronic modes \cite{Fidler2010}, double-exciton resonances \cite{Bruggemann2011,Bubilaitis2019,Dostal2018} or specific exciton relaxation pathways \cite{Zhang2015,Li2013} have been revealed.
To collect the complete information, five excitation pulses are needed for generation
of interaction pathway-specific detailed picture. The full set of
information then comes up in five dimensions as five independent time
intervals between interactions should be considered. 
Utilizing e. g. the gradient assisted photon echo (GRAPE)
methodology, the fifth-order nonlinear polarization  through three sequential single-quantum coherences have been  measured in
parallel \cite{Fidler2010} and  presumably  could be extended for 5D measurement.
However, the
information contents in the corresponding full-scale experiments becomes
huge, and interpretation becomes challenging.

The approach presented in this paper is not related to extracting
all possible information at a certain order, but, instead, is specifically
geared to extract only small amount of information, but with a minimal
amount of effort with high accuracy. A similar approach has been utilized to gain 
high-order electronic correlations in quantum wells \cite{Turner2010}.

\begin{figure}
\begin{centering}
\includegraphics[width=0.5\textwidth]{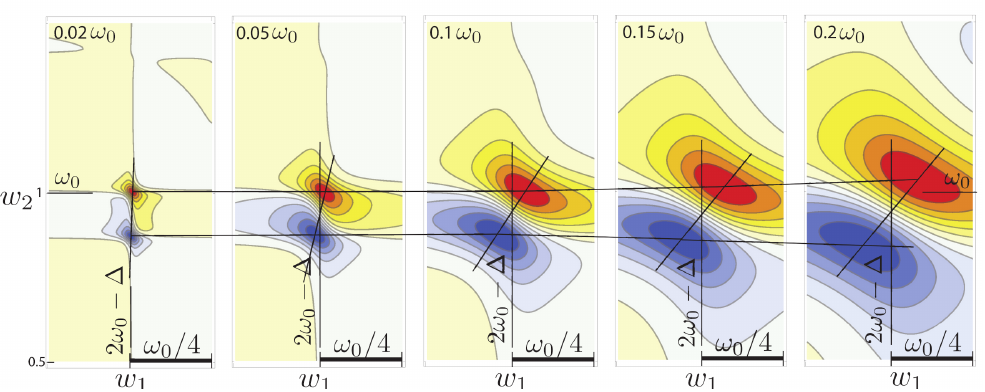}
\par\end{centering}
\caption{\label{fig:regarding-anharmonicity-1}Numerical simulations of simple
2DIR signal of the Morse oscillator with the increasing linewidth
parameter $\gamma$ shown in insets of the subplots.}
\end{figure}

This specificity  can be  demonstrated with respect to the vibrational
anharmonicity that is often targeted in 2DIR of proteins. Simple numerical
simulations in Fig. \ref{fig:regarding-anharmonicity-1} demonstrate
that the anharmonicity that is being extracted from simple 2DIR measurement
may be distorted by large broadening, and thus, cannot be used to
accurately determine the corresponding hamiltonian parameters. Meanwhile
Fig. \ref{fig:quantum-broad-classical}(top) demonstrates that the
anharmonicity is revealed not only on vertical but also on horizontal
axes and can be accessed without interference from line overlaps.

It should be noted that other anharmonic properties should be included
when using more accurate simulation approaches. For example, 0-2,
0-3 etc. transition dipole elements of the Morse potential are not
zero, additionally at high excitation levels, continuums states can
be reached. On the spectroscopy side this could indicate an interesting
turn for the spectroscopy analysis, while on the chemical side this
can be used to initial chemical reactions or molecular isomerization.

The approach described in this paper in general applies not only to
vibrational excitations but also to electronic states of molecules
or atoms, but would require careful analysis of the spectra as electronic
oscillators are highly anharmonic and positions of double-triple quanta
excitations are poorly \emph{a priori} anticipated (that may not be
the case for Rydberg atoms in gas phase, etc.)

The other question that comes up is what kind of new information can
be extracted from such multi-dimensional signals. 
In electronic time
resolved spectroscopy the excitation relaxation pathways are primary
targets and, for example, according to the perturbation theory the
relaxation process depends on the two point fluctuation correlation
function \cite{val-ab-knyga}. In principle the $n$-th order nonlinear
signal is dependent on the $n+1$ point correlation function \cite{mukamelbook}.
So $1$-st order response (linear absorption) is related to the two-point
correlation function $C(t_{2},t_{1})$ which reduces to $C(t_{2}-t_{1})$
due to the assumption of ergodicity (equilibrium before the measurement),
and is in principle sufficient to reveal the relaxation phenomena.
Indeed the dephasing and relaxation processes determined spectral
lineshape positions and broadening. The correlation function here
is considered in a broad sense to include also the electronic degrees
of freedom. The third order response function maps the four-point
correlation functions. The electronic degrees of freedom (i. e. energy
level structure) become reflected onto the 2D plot in two-dimensional
coherent spectroscopy, while the vibrational part of the correlation
function results in spectral lineshapes. However, assuming the fluctuation
process as a Gaussian process, all multi-point correlation functions
split into a set of simple two-point, $C(t_{i}-t_{j})$, correlation
functions. Consequently, going to higher orders does not necessarily
generate novel information. Recently at fifth order to the field the
exciton-exciton annihilation processes have  been considered that
encapsulate the novel four particle correlation properties as they
originate from $\hat{a}^{\dagger}\hat{a}^{\dagger}\hat{a}\hat{a}$
terms \cite{Bruggemann2011,Bubilaitis2019,Dostal2018}, however, focusing on exciton
migration again restricts the focus again only to the one-particle
diffusion problem.

The level of complexity becomes overwhelming in biological  systems where  interactions
between electronic and vibrational degrees of freedom may
have important implications for rapid and efficient energy
transfer. Development of higher-order spectroscopies that may enable
unambiguous assignment of signals to specific pathways is very active.
Highly nonlinear Raman approaches have been proposed since 2000 \cite{Tanimura2000}. 2D Resonance Raman techniques denoted by femtosecond Stimulated Raman Spectroscopy (FSRS), yield a subset of the quantum pathways \cite{Kubarych2003,Molesky2014,Molesky2015}
reveal electronic and vibrational interactions, however, suffer from contributions from cascades of third-order processes \cite{Kirkwood2000,Cho2000,Mehlenbacher2009}.
Direct
correlation between impulsively driven low-frequency modes
such as phonons, vibrations and (multi-)excitons with quantum
coherence selectivity through control of resonance can specifically be probed by fifth order   gradient-assisted
multidimensional electronic Raman spectroscopy (GAMERS)  free from low order cascades \cite{Spencer2017}.
Hence, target-specific approaches are advantageous.

It should be noted that large field intensities are often necessary
to reach good signal to noise ratio at high nonlinearity orders. Then
high order nonlinearities become involved in apparent low degree phase
matching configurations that are supposed to be sensitive only to
low orders to the field. For example the traditional photon echo $k_{s}=-k_{1}+k_{2}+k_{3}$
configuration may be plagued by other superpositions, like $k_{s}=-k_{1}-k_{1}+k_{1}+k_{2}+k_{3}$,
etc., which come from the fifth order \cite{Dostal2018,Bubilaitis2019}.
Similarly, seventh, ninth and so on orders contribute to the specific
phase matching configuration. Therefore studying high-order nonlinearities
in seemingly low order phase matching configurations, e. g. exciton
annihilation in pump-probe, carries large nonzero background. Instead
in our suggested approach we have a zero background signal, what allows
to cleanly identify the specific order without \textquotedbl poisoning\textquotedbl{}
by higher orders when lowest possible excitation intensity is maintained.
Alternatively, low order cascading signals may obscure the highly nonlinear signals in simple phase matching configurations.
Approach suggested in this paper could be free from cascading contributions if spatial phase matching in non-collinear laser configuration is applied. In collinear scheme with phase matching the error control must be very high level to avoid cascading. 

Concluding, a simple two-exitation-pulse nonlinear spectroscopy approach
is proposed to reveal quantum ladder of states in an anharmonic oscillator.
Consequently this could be employed in reconstruction of whole potential
surface at a desired excitation level.

%


\section*{Acknowledgements}

The author acknowledges financial support of Lithuanian Science Council (grant No: SMIP-20-47)

\end{document}